\documentclass[conference]{IEEEtran}
\IEEEoverridecommandlockouts
\usepackage{cite}
\usepackage{balance}
\usepackage{float}
\usepackage{amsmath,amssymb,amsfonts}
\usepackage{algorithmic}
\usepackage{graphicx}
\usepackage{textcomp}
\usepackage{xcolor}
\usepackage{acronym}
\usepackage{subcaption}
\usepackage[hidelinks]{hyperref}

\acrodef{vlc}[VLC]{visible light communication}
\acrodef{vlp}[VLP]{visible light positioning}
\acrodef{lifi}[LiFi]{light fidelity} 
\acrodef{iot}[IoT]{Internet-of-Things}
\acrodef{rf}[RF]{radio frequency}
\acrodef{gbps}[Gbps]{gigabit-per-second}
\acrodef{mm wave}[mm-Wave]{millimetre wave}
\acrodef{thz}[THz]{terahertz}
\acrodef{irs}[IRS]{intelligent reflecting surfaces}
\acrodef{snr}[SNR]{signal-to-noise ratio}
\acrodef{em wave}[EM Wave]{electromagnetic wave}
\acrodef{cs}[Cs]{secrecy capacity}
\acrodef{pls}[PLS]{physical layer security}
\acrodef{ofdm}[OFDM]{orthogonal frequency division multiplexing}
\acrodef{ma}[MA]{multiple access}
\acrodef{tdma}[TDMA]{time division multiple access}
\acrodef{ofdma}[OFDMA]{orthogonal division multiple access}
\acrodef{noma}[NOMA]{non-Orthogonal multiple access}
\acrodef{cagr}[CAGR]{compound annual growth rate}
\acrodef{m2m}[M2M]{machine-to-machine}
\acrodef{mimo}[MIMO]{multiple-input and multiple-output}
\acrodef{owc}[OWC]{optical wireless communication}
\acrodef{uv}[UV]{ultraviolet}
\acrodef{ir}[IR]{infrared}
\acrodef{fso}[FSO]{free-space optical}
\acrodef{uv}[UV]{ultraviolet}
\acrodef{uwc}[UWC]{underwater communication}
\acrodef{vl}[VL]{visible light}
\acrodef{nm}[nm]{nano-meters}
\acrodef{led}[LEDs]{light-emitting-diodes}
\acrodef{ld}[LDs]{laser-diodes}
\acrodef{wifi}[WiFi]{wireless fidelity}
\acrodef{ap}[AP]{access point}
\acrodef{ue}[UE]{user equipment}
\acrodef{fov}[FoV]{field-of-view }
\acrodef{a/w}[A/W]{ampere per watt}
\acrodef{im-dd}[IM-DD]{intensity modulation and direct detection }
\acrodef{los}[LoS]{line-of-sight}
\acrodef{nlos}[NLoS]{non line-of-sight}
\acrodef{pin}[PIN]{positive-intrinsic-negative }
\acrodef{mems}[MEMS]{micro-electro-mechanical systems}
\acrodef{isi}[ISI]{inter-symbol interference}
\acrodef{re}[REs]{reflecting elements}
\acrodef{ber}[BER]{bit error rate}
\acrodef{fpa}[FPA]{fixed power allocation}
\acrodef{cs}[Cs]{secrecy capacity}
\acrodef{dco}[DCO]{direct current-biased optical}
\acrodef{ml}[ML]{machine learning}

\acrodef{pls}[PLS]{physical layer security}
\acrodef{ga}[GA]{genetic algorithm}
\acrodef{dco-ofdm}[DCO-OFDM]{direct current optical frequency division multiplexing }
\acrodef{led}[LED]{light-emitting diode}
\acrodef{los}[LoS]{line-of-sight}
\acrodef{nlos}[NLoS]{non-line-of-sight}
\acrodef{pd}[PD]{photo-detector}
\acrodef{cir}[CIR]{channel impulse response}
\acrodef{cfr}[CFR]{channel frequency response}
\acrodef{isi}[ISI]{intersymbol interference}
\acrodef{snr}[SNR]{signal-to-noise ratio}
\acrodef{es}[ES]{exhaustive search}
\acrodef{ar}[AR]{augmented reality}
\acrodef{vr}[VR]{virtual reality}
\acrodef{vcsel}[VCSEL]{vertical-cavity surface-emitting laser}
\acrodef{rss}[RSS]{received signal strength}
\acrodef{aoi}[AoI]{angle of incidence}
\acrodef{aoa}[AoA]{angle of arrival}
\acrodef{tdoa}[TDoA]{time difference of arrival}
\acrodef{toa}[ToA]{time of arrival}
\acrodef{dnn}[DNN]{deep neural network}
\acrodef{cdf}[CDF]{cumulative distribution function}
\acrodef{crlb}[CRLB]{Cramér–Rao lower bound}
\def\BibTeX{{\rm B\kern-.05em{\sc i\kern-.025em b}\kern-.08em
    T\kern-.1667em\lower.7ex\hbox{E}\kern-.125emX}}
\begin{document}

\title{A Scanning-Based Indoor Optical Wireless Positioning System with Single VCSEL\\}

\author{\IEEEauthorblockN{Yicheng Dong, Rashid Iqbal, Julien Le Kernec, Hanaa Abumarshoud}
\IEEEauthorblockA{James Watt School of Engineering, University of Glasgow, Glasgow, G12 8QQ, UK}
\IEEEauthorblockA{e-mail: \{y.dong.2, r.iqbal.1\}@research.gla.ac.uk, \{Julien.LeKernec, Hanaa.Abumarshoud\}@glasgow.ac.uk}

}

\maketitle

\begin{abstract}
This paper presents a novel indoor visible light positioning (VLP) system utilising one vertical-cavity surface-emitting laser installed at the ceiling centre of a space. The system offers three-dimensional localisation by sweeping through space at one-degree resolution in two dimensions (azimuth and elevation), significantly simplifying hardware. Through incorporating the angle of arrival and received signal strength, this system demonstrates excellent precision in indoor positioning. Simulation results verify that the system attains sub-centimetre precision for most test points, outperforming conventional multi-transmitter VLP schemes in cost-efficiency and simplicity.
\end{abstract}


\section{Introduction}
The growth of \ac{iot} devices over recent years has rendered the \ac{rf} spectrum highly crowded and so in need of fresh solutions. At the same time, high-speed wireless applications have emerged in various industries like healthcare, smart homes, virtual reality, and augmented reality. The applications require very high data rates that traditional wireless technologies cannot support\cite{saad1}. That is where the fifth-generation cellular network (5G) comes in and will enable \ac{gbps} data rates, increased reliability, and massive connectivity with reduction of costs and power consumption. However, introduction of beyond's 5G cellular network (B5G) has problems associated with it, especially at \ac{mm wave} and \ac{thz} frequencies famous for their high penetration attenuations \cite{Bariah1,Islam1}. 

To mitigate such problems, \ac{vlc} is taken into consideration as one of the solutions. \Ac{vlc} operates from 400 THz to 790 \ac{thz} and offers various advantages such as multi-\ac{gbps} wireless transmission, large data density, relief from interference, beam steering, ease in integration, lighting feature, and cost-effectiveness \cite{Haas}. Among these benefits, \ac{vlc} has limited communication distance and loss of signal in obstructed environments \cite{Hanaa,9524909}.

To fully exploit these advantages in practical scenarios, especially indoors, accurate user positioning becomes a critical requirement. In VLC systems, the directionality and narrow beam footprint of light-based transmitters mean that signal quality and availability are highly dependent on the user’s exact location and orientation. Without precise positioning, users may fall outside the beam coverage, leading to degraded communication performance or complete signal loss. Moreover, enabling advanced features such as beam steering, handover, and user-centric resource allocation demands real-time knowledge of user location. Therefore, designing robust and efficient positioning mechanisms is essential for unlocking the full potential of VLC-based communication systems in smart indoor environments \cite{10195985,abumarshoud2020optical,9614037}.

To this end, recent research has gained attention for using \ac{vlc} not only for high-speed data transmission but also for accurate user positioning. Traditionally, VLC-based localisation has relied on \acp{led}, utilising signal features such as \ac{rss}, \ac{aoa}, and \ac{toa} to estimate user positions in indoor settings. However, LEDs generally emit wide-angle light, which limits spatial resolution and positioning precision. As an alternative, \acp{vcsel} offer compact, low-power, and narrow-beam optical sources that enable high-resolution spatial sensing. Due to their Gaussian beam profile and ability to support multi-mode diversity, VCSELs are particularly well-suited for precise indoor positioning. Building on this, recent studies have applied \ac{ml} techniques to leverage signal characteristics from VCSEL-based systems. For instance, the work in \cite{prob2} demonstrates that a \ac{dnn} can successfully estimate the 3D position and orientation. Moreover, the work in \cite{iqbal} uses duel mode \ac{vcsel} to estimate the user position and orientation.

In this work, we present an indoor \ac{vlp} system that utilises a single \ac{vcsel} mounted at the centre of the ceiling in an enclosed environment. By sweeping the beam at one-degree resolution in both azimuth and elevation, the system achieves three-dimensional localisation while significantly simplifying the required hardware. We further validate the proposed architecture through simulations, demonstrating that it achieves sub-centimetre positioning accuracy across most test points. Compared to conventional multi-transmitter VLP schemes, our approach offers improved cost-efficiency and reduced complexity without compromising accuracy.

\section{System Model and proposed approach}
\begin{figure*}[!ht]
\centering
\includegraphics[width=\textwidth]{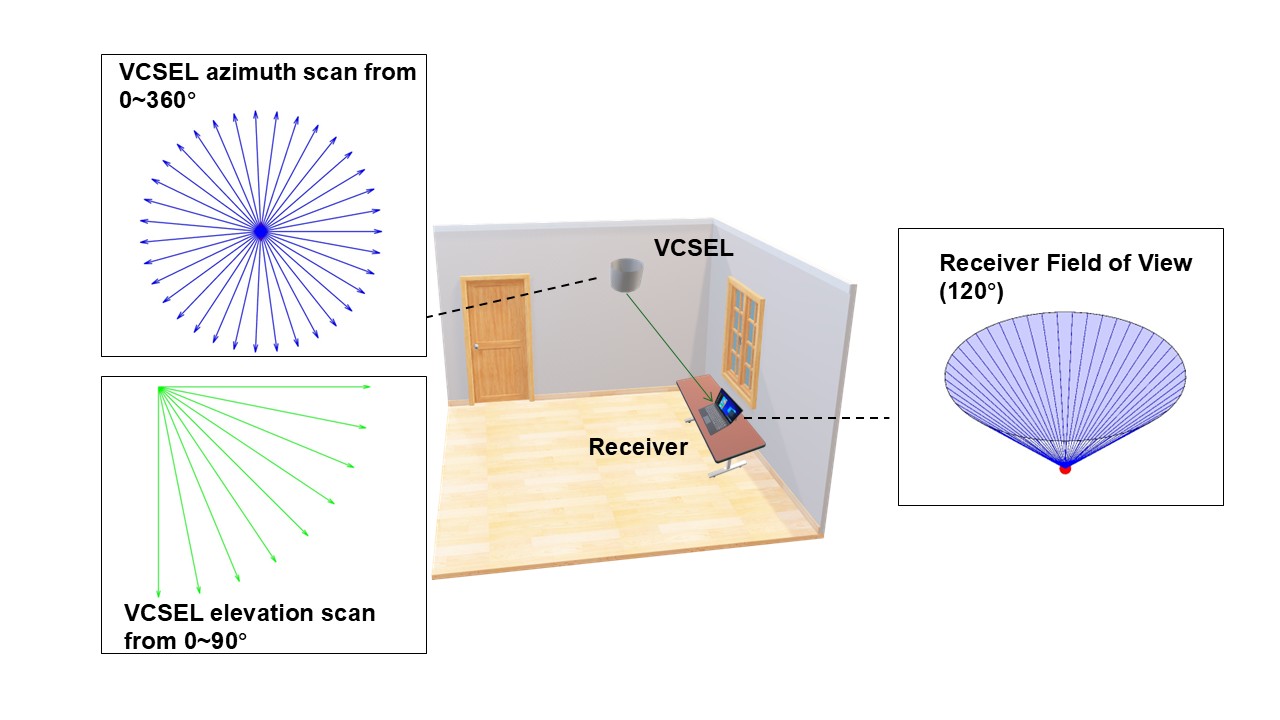}
\caption{System model of an indoor single-\ac{vcsel} \ac{vlp} system with one user}
\label{room}
\end{figure*}

We propose an indoor positioning system leveraging a single \ac{vcsel} strategically placed at the room's ceiling centre as shown in Fig. \ref{room}. The receiver is assumed to have a \ac{fov} where the incident signal beyond this field would not be received.

\begin{figure}[!ht]
    \centering
    \includegraphics[width=\linewidth]{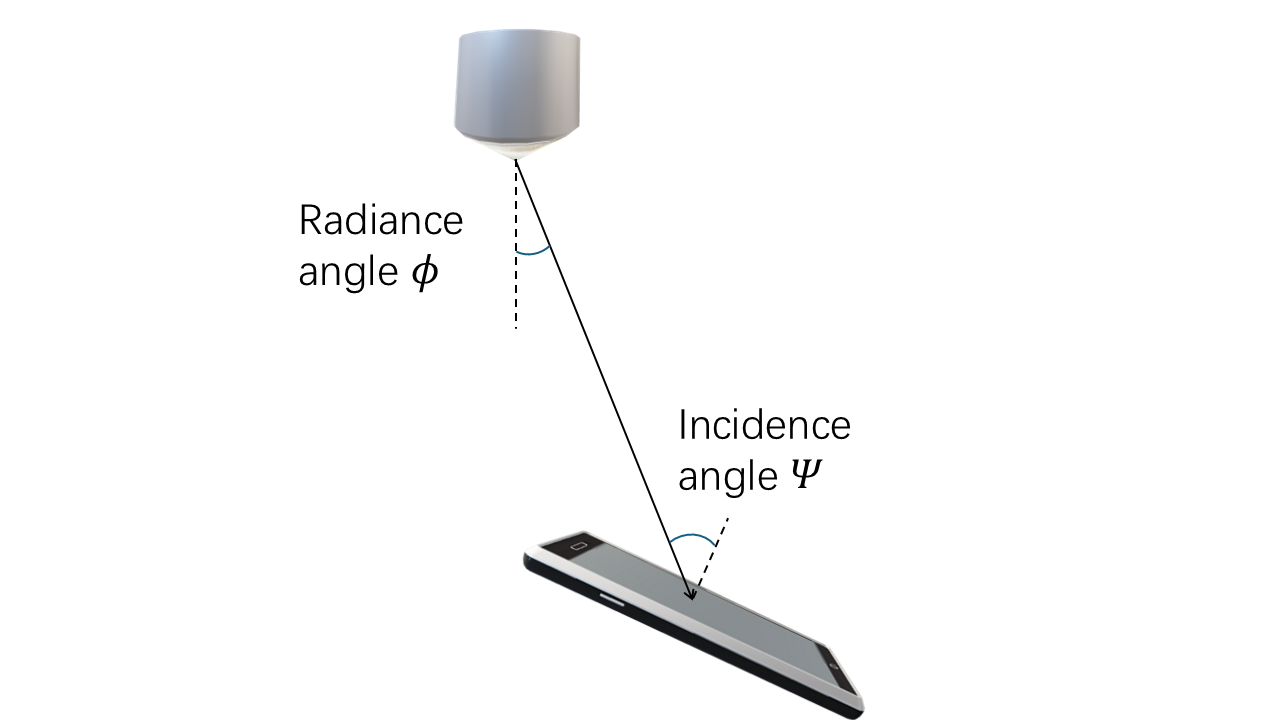}
    \caption{The angles of the positioning system}
    \label{angles}
\end{figure}

\begin{figure}[!ht]
    \centering
    \includegraphics[width=\linewidth]{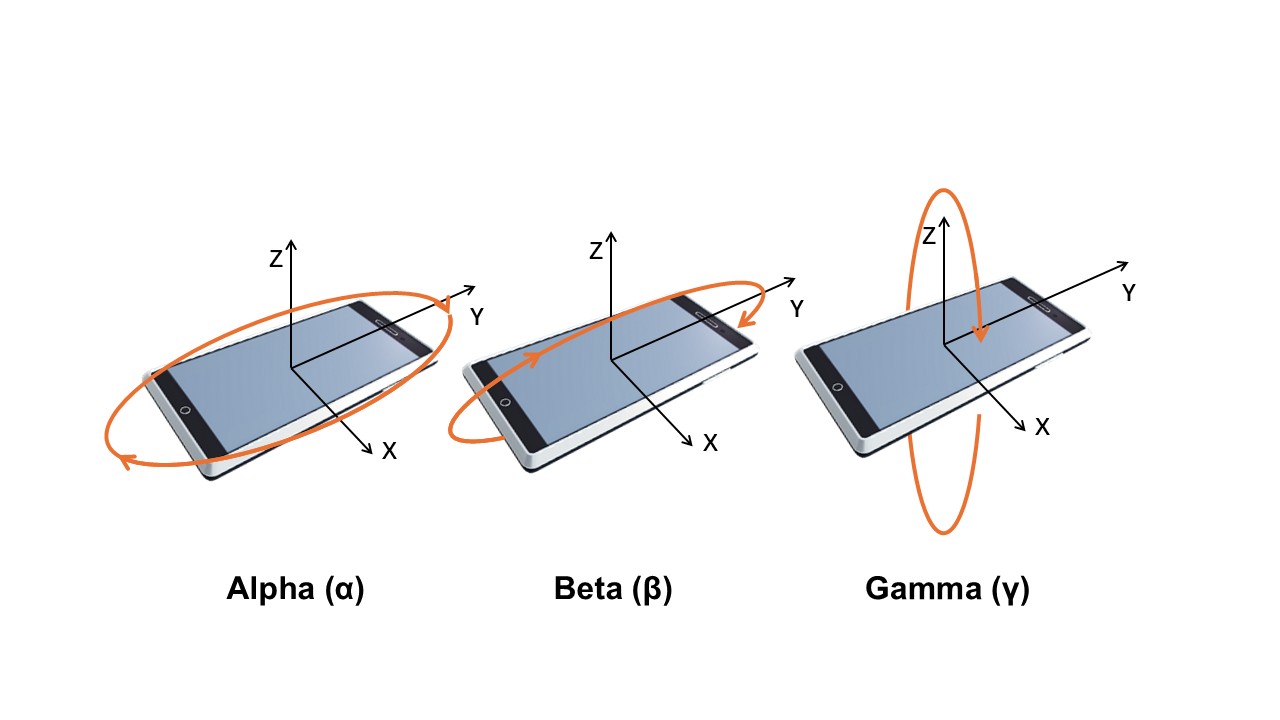}
    \caption{A demonstration of the user device orientation angles}
    \label{fig:Orientatioin angles}
\end{figure}

To determine the distance between the \ac{vcsel} and the user, the received optical power is used, which is calculated as
\begin{equation}
P_r = I(d, \phi) \cdot A_{PD} \cdot \cos(\psi) + N_p,
\label{eq_pr}
\end{equation}

where $N_p$ is the simulated noise level, and is calculated as\cite{Kazemi_2025}
\begin{equation}
    N_p = \mathcal{N}/B,
\end{equation}

where $\mathcal{N}$ denotes the noise variance and B denotes the system bandwidth. Moreover, $A_{PD}$ is the \ac{pd} active area, the cosine of the incidence angle is given by
\begin{equation}
\cos(\psi) = \frac{\mathbf{d} \cdot \mathbf{n}_{PD}}{\|\mathbf{d}\|}, 
\end{equation}

where ${\|\mathbf{d}\|}$ is the Euclidean distance between the VCSEL and the user and $n_{PD}$ is the normal vector of the user device which is calculated as
\begin{equation}
    \mathbf{n}_{\text{PD}} = 
    \begin{bmatrix}
        \cos\gamma \sin\alpha \sin\beta + \cos\alpha \sin\gamma \\
        \sin\alpha \sin\gamma - \cos\alpha \cos\gamma \sin\beta \\
        \cos\gamma \cos\beta
    \end{bmatrix},
\end{equation}
where $\alpha$, $\beta$, and $\gamma$ represent the roll, pitch, and yaw angles,respectively, as shown in Fig \ref{fig:Orientatioin angles}.

Furthermore, $I(d, \phi)$ at distance $d$ and the radiance angle $\phi$ is expressed as
\begin{equation}
I(d, \phi) = \frac{2P_{opt}}{\pi w(d)^2} \exp\left(-\frac{2d^2 \sin^2(\phi)}{w(d)^2}\right),
\end{equation}
with $P_{opt}$ representing transmitted power and beam radius $w(d)$ calculated by
\begin{equation}
w(d) = w_0 \sqrt{1 + \left(\frac{\lambda d \cos(\phi)}{\pi w_0^2}\right)^2},
\end{equation}
where $w_0$ is the initial beam waist radius and $\lambda$ is the wavelength of the VCSEL.

Since the beam divergence is sufficiently narrow and that there is precise alignment between the transmitter and the receiver, the angle $\phi$ between their axes can be considered negligible. Under this assumption, the $\phi = 0$ and the equation (\ref{eq_pr}) is simplified as

\begin{equation}
P_r = \frac{2 P_{\text{opt}} \cdot A_{\text{PD}} \cdot \cos(\psi)}{\pi w_0^2 \left( 1 + \left( \frac{\lambda d}{\pi w_0^2} \right)^2 \right)} + N_p.
\label{eq_prfinal}
\end{equation}

\subsection{Synchronisation of the system} 
\label{sync}

In the proposed scanning scheme, the \ac{vcsel} sequentially emits beams in different directions, and the receiver records the corresponding signal strength at each time step. For accurate position estimation, it is essential that each received sample is correctly associated with its corresponding transmitted beam. This requirement is referred to as time synchronisation \cite{proakis2001digital}.

Formally, let each beam dwell for $T_{\mathrm{step}}$ seconds. If the timing offset between the transmitter and the receiver is smaller than half of this duration, i.e. $|\Delta t| < \tfrac{1}{2} T_{\mathrm{step}},$ then the receiver samples can be reliably matched to the correct beam indices. In this synchronised case, the $j$-th entry of the measurement vector $\mathbf{y}$ corresponds directly to the $j$-th column of the beam-direction matrix $\mathbf{U}$. As a result, the beam producing the maximum received power can be uniquely identified, and its associated incidence vector is used for position estimation.

If the system is not synchronised, the index mapping between transmitted beams and received samples is lost, which leads to selecting an incorrect incidence vector and hence large positioning errors. To avoid this, synchronisation can be ensured either by hardware-level clock alignment or by embedding known pilot signals in the transmission and applying a cross-correlation procedure at the receiver to realign the indices.

\subsection{Synchronised Beam–Measurement Matching and Position Estimation}
\label{est}

As is mentioned above, a timed scan is implemented where the VCSEL emits \(M\) beams in sequence and the receiver captures \(M\) corresponding samples. The following compact matrix formulation describes how the beam unit vectors and measured intensities are matched and how the user position is estimated.

The unit beam vectors can be expressed as columns of the \(3\times M\) matrix
\[
\mathbf{U} = \big[\,\hat{\mathbf{u}}_1 \;\; \hat{\mathbf{u}}_2 \;\; \cdots \;\; \hat{\mathbf{u}}_M\,\big]
\in\mathbb{R}^{3\times M},
\qquad
\hat{\mathbf{u}}_j=(u_{x,j},u_{y,j},u_{z,j})^\top.
\]
And the receiver's measured received powers $y_j$ can be written as
\[
\mathbf{y} = \begin{bmatrix} y_1 \\ y_2 \\ \vdots \\ y_M \end{bmatrix}
\in\mathbb{R}^{M},
\qquad
y_j = P^{(j)} + N_p,
\]
where \(P^{(j)}\) is the noise-free received power for beam \(j\).

The predicted received power for beam \(j\) at distance \(d\) is
\begin{equation}
P_j(d,\psi_j)
\;=\;
\frac{2P_{\mathrm{opt}} A_{PD} \cos\psi_j}
{\pi\!\left(w_0^2 + \dfrac{\lambda^2 d^2}{\pi^2 w_0^2}\right)},
\label{eq:gj}
\end{equation}
where \(\psi_j\) is the incidence angle for beam \(j\). Combined with the received power matrix
\[
\mathbf{P}(d) = \begin{bmatrix} P_1(d,\psi_1) \\ P_2(d,\psi_2) \\ \vdots \\ P_M(d,\psi_M) \end{bmatrix}.
\]

The measurement model is therefore
\begin{equation}
    \mathbf{y} = \mathbf{P}(d) + N_p.
\end{equation}

We assume time synchronisation such that the receiver sample index \(j\) corresponds to transmit beam \(j\). Under this assumption the system identifies the beam that points to the user by selecting the largest measured intensity
\begin{equation}
\hat{k} \;=\; \arg\max_{j\in\{1,\dots,M\}} \; y_j.
\label{eq:select}
\end{equation}
The synchronisation guarantee ensures the measurement index equals the transmit index, so \(\hat{k}\) corresponds to the candidate beam vector
\begin{equation}
    \hat{\mathbf{u}}_{\hat{k}} \;=\; \mathbf{U}\,\mathbf{e}_{\hat{k}},
\end{equation}

where \(\mathbf{e}_{\hat{k}}\) is the \(\hat{k}\)-th canonical basis vector.

Using the selected measurement \(y_{\hat{k}}\) as the observed received power for beam \(\hat{k}\), distance \(\hat d\) can be estimated by inverting \eqref{eq:gj} with \(P_r = y_{\hat{k}}\),
\begin{equation}
\hat d \;=\; \sqrt{ \frac{\pi w_0^2}{y_{\hat{k}} \lambda^2} \Big( 2 A_{PD} P_{\mathrm{opt}} \cos\psi_{{k}} - y_{\hat{k}} \pi w_0^2 \Big) },
\label{eq:dhat}
\end{equation}

Using \eqref{eq:dhat} and the \ac{vcsel} position \(\mathbf{p}_s\). The estimated user position is calculated as 
\begin{equation}
{\; \hat{\mathbf{p}} \;=\; \mathbf{p}_s \;+\; \hat d \; \hat{\mathbf{u}}_{\hat{k}} \;=\; \mathbf{p}_s + \hat d \,\mathbf{U}\,\mathbf{e}_{\hat{k}} \; }.
\label{eq:phat}
\end{equation}

Considering \(\mathbf{p}\) is the true user position, the position error would be
\begin{equation}
    e \;=\; \| \mathbf{p} - \hat{\mathbf{p}} \|.
\end{equation}

\subsection{User Orientation}

To model the natural variations in user device orientation, each orientation angle is treated as a random variable drawn from a Laplace distribution \cite{orientation}. The Laplace distribution is a two-parameter continuous distribution with a probability density function given by
\begin{equation}
    f(x \mid \mu, b) = \frac{1}{2b} \exp\left(-\frac{|x - \mu|}{b}\right)
\end{equation}
where \( \mu \) is the mean and \( b \) is the scale parameter. The scale parameter is related to the standard deviation \( \sigma \) by
\begin{equation}
    b = \frac{\sigma}{\sqrt{2}}
\end{equation}
Orientation samples are drawn using the inverse transform method as
\begin{equation}
    r = \mu - b \cdot \operatorname{sign}(u) \cdot \ln(1 - 2|u|)
\end{equation}
where \( u \sim \mathcal{U}(-0.5, 0.5) \) is a uniform random variable. Using this approach, both the azimuth \( \phi \) and elevation \( \theta \) angles are independently sampled as
\begin{equation}
    \phi = \mu_{\phi} - b_{\phi} \cdot \operatorname{sign}(u_1) \cdot \ln(1 - 2|u_1|), \quad u_1 \sim \mathcal{U}(-0.5, 0.5)
\end{equation}
\begin{equation}
    \theta = \mu_{\theta} - b_{\theta} \cdot \operatorname{sign}(u_2) \cdot \ln(1 - 2|u_2|), \quad u_2 \sim \mathcal{U}(-0.5, 0.5)
\end{equation}
Finally, the orientation vector of the photodetector is computed from these angles using a spherical-to-Cartesian transformation
\begin{equation}
    \mathbf{n}_{p_D} = \begin{bmatrix}
        \cos(\theta)\cos(\phi) \\
        \cos(\theta)\sin(\phi) \\
        \sin(\theta)
    \end{bmatrix}
\end{equation}
This formulation allows the orientation of the receiver to vary realistically during simulations while capturing both typical fluctuations and occasional extreme deviations.

We adopt a rotating scheme that enables the \ac{vcsel} to scan the indoor environment with one-degree angular resolution in both azimuth $0^\circ$–$360^\circ$ and elevation $0^\circ$–$90^\circ$, enabling comprehensive coverage with minimal infrastructure complexity. It is worth noting that according to the research of \cite{lidar}, similar scanning devices such as lidars are sufficiently efficient, enabling the scanning process to be conducted within a second. The localisation methodology integrates angular and distance measurements. We assume that the \ac{vcsel} rotation is synchronised with the receiver, allowing the receiver to identify the precise angle of radiance corresponding to peak \ac{rss}. This synchronisation ensures accurate determination of the incidence angle at the receiver location. After the \ac{rss} and all required angles are achieved, the distance is calculated using the reverse of equation (\ref{eq_prfinal}). It is worth noting that a similar rotating concept was applied in \cite{PD_rotation} for the purpose of increasing receiver coverage.

\section{Simulation results}

{\small \begin{table}[!ht]
\hspace*{-0.04in}
\caption{Simulation Parameters}
\label{table:simulation_parameters}
\centering
\begin{tabular}{|c|c|}
\hline
\textbf{Parameters} & \textbf{Values} \\
\hline
Transmitted power [Watt] & \( P_{\scriptscriptstyle opt} = 0.001\) \\
\hline
System bandwidth [GHz] & \( B = 2 \) \\
\hline
PD physical area [cm\(^2\)] & \( A_{\scriptscriptstyle pd} = 1 \) \\
\hline
PD \ac{fov} & \( \Psi_{\scriptscriptstyle \text{FoV}} = 120^\circ \) \\
\hline
Noise variance &  \(\mathcal{N} = 5*10^{-14}\) \\
\hline
Beam wavelength [nm] & \( \lambda = 950\) \\
\hline
Beam waist radius [$\mu$m] & \(\omega_0 = 5.9 \) \\
\hline
Maximum user height [m] & \( H_{max} = 2.5 \) \\
\hline
Minimum user height [m] & \( H_{min} = 0 \) \\
\hline
Mean orientation angle $\alpha$ & \(\mu_{\alpha} = 0^\circ\) \\
\hline
Standard deviation of $\alpha$ & \(\sigma_{\alpha} = 10\) \\
\hline
Mean orientation angle $\beta$ & \(\mu_{\beta} = 0^\circ\) \\
\hline
Standard deviation of $\beta$ & \(\sigma_{\beta} = 30\) \\
\hline
Mean orientation angle $\gamma$ & \(\mu_{\gamma} = 0^\circ\) \\
\hline
Standard deviation of $\gamma$ & \(\sigma_{\gamma} = 10\) \\
\hline
Scanning times $M$ & \( M = 90*360=32400 \) \\
\hline

\end{tabular}
\label{table 1}
\end{table}}

In this section, we discuss the simulation results of the positioning system of an indoor room with dimensions of $1m\times1m\times3m$. Unless otherwise specified, the simulation parameters are shown in Table \ref{table:simulation_parameters}. The main goal of these simulations is to demonstrate the effectiveness of the proposed positioning system for the indoor scenario.

\begin{figure*}[t]
\centering

\begin{minipage}[t]{0.49\textwidth}
    \centering
    \includegraphics[width=\linewidth]{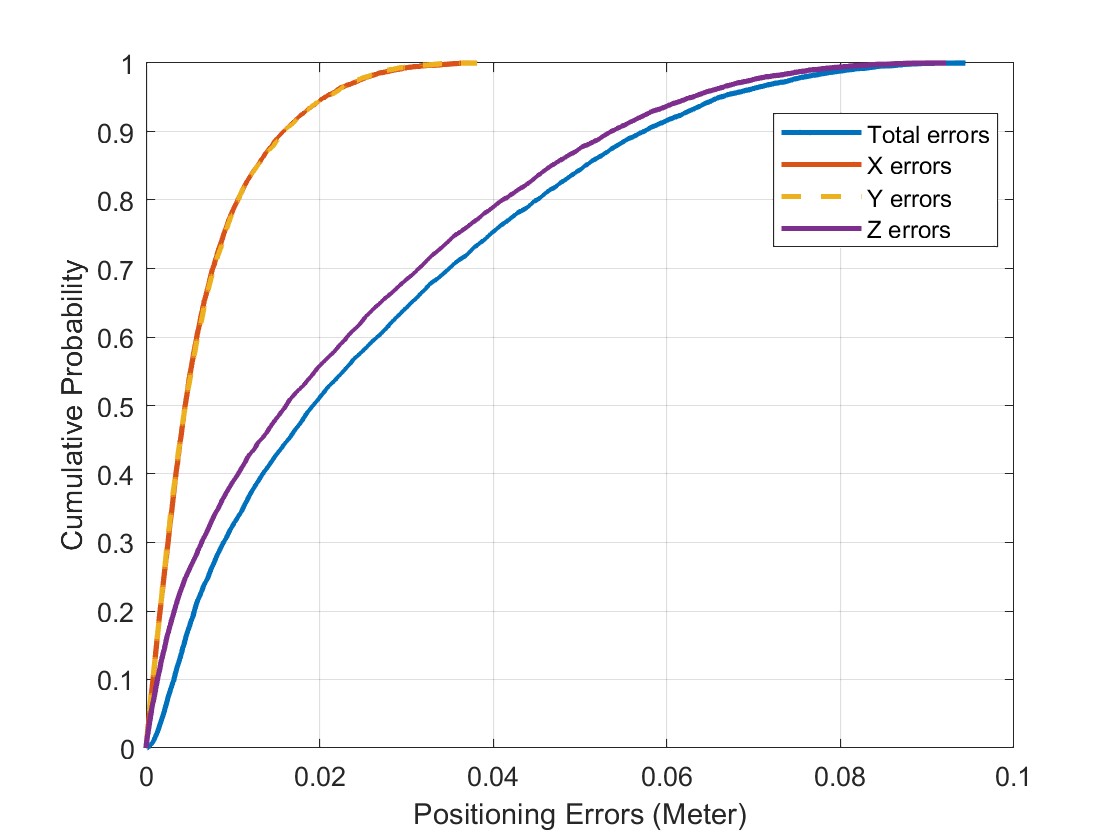}
    \caption{CDF of position error with fixed receiver orientation.}
    \label{fig:cdf fix}
\end{minipage}\hfill
\begin{minipage}[t]{0.49\textwidth}
    \centering
    \includegraphics[width=\linewidth]{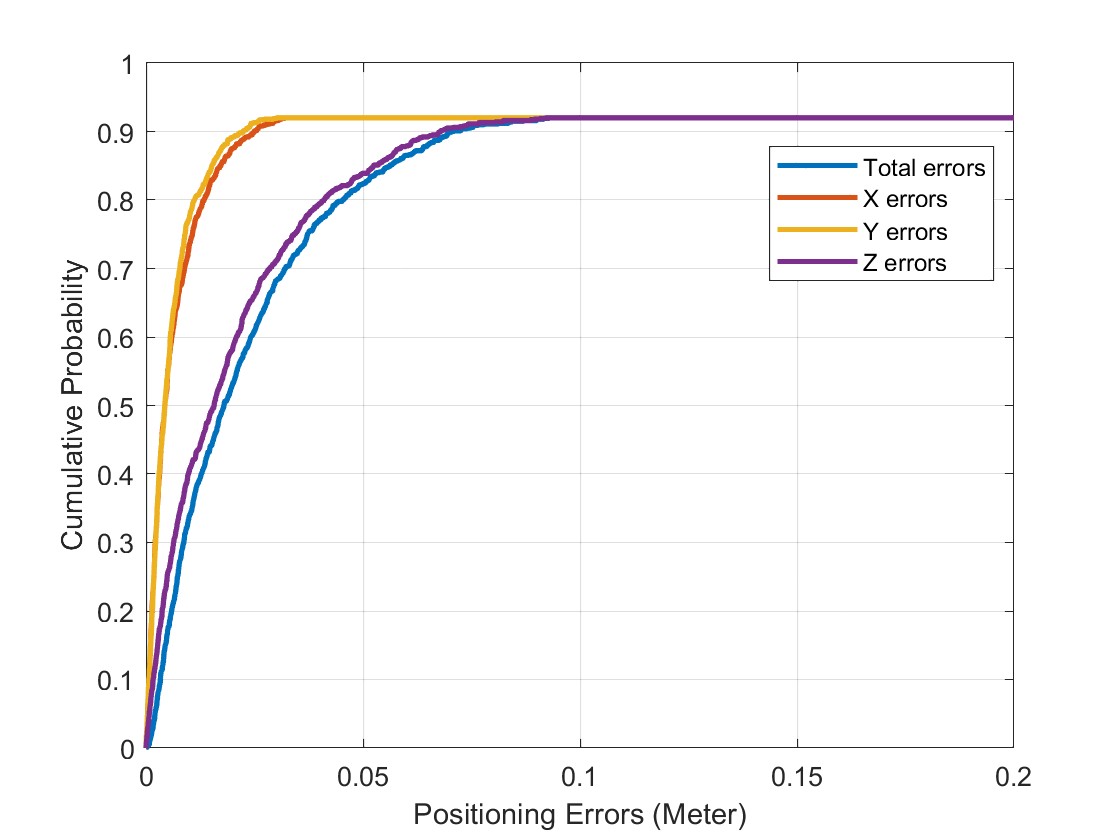}
    \caption{CDF of position error with random receiver orientation.}
    \label{fig:cdf rand}
\end{minipage}
\end{figure*}

\begin{figure*}
\begin{minipage}[t]{0.49\textwidth}
    \centering
    \includegraphics[width=\linewidth]{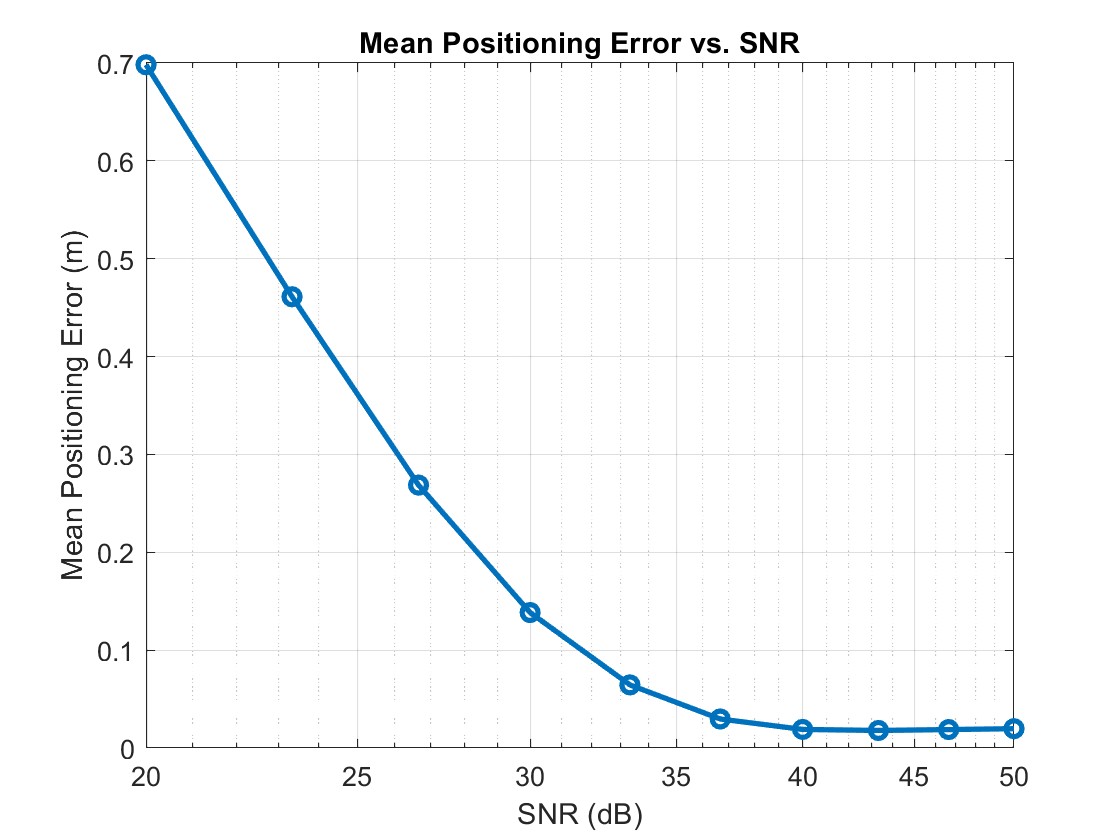}
    \caption{Mean positioning error vs \ac{snr} under fixed orientation.}
    \label{fig:mean error fix}
\end{minipage}\hfill
\begin{minipage}[t]{0.49\textwidth}
    \centering
    \includegraphics[width=\linewidth]{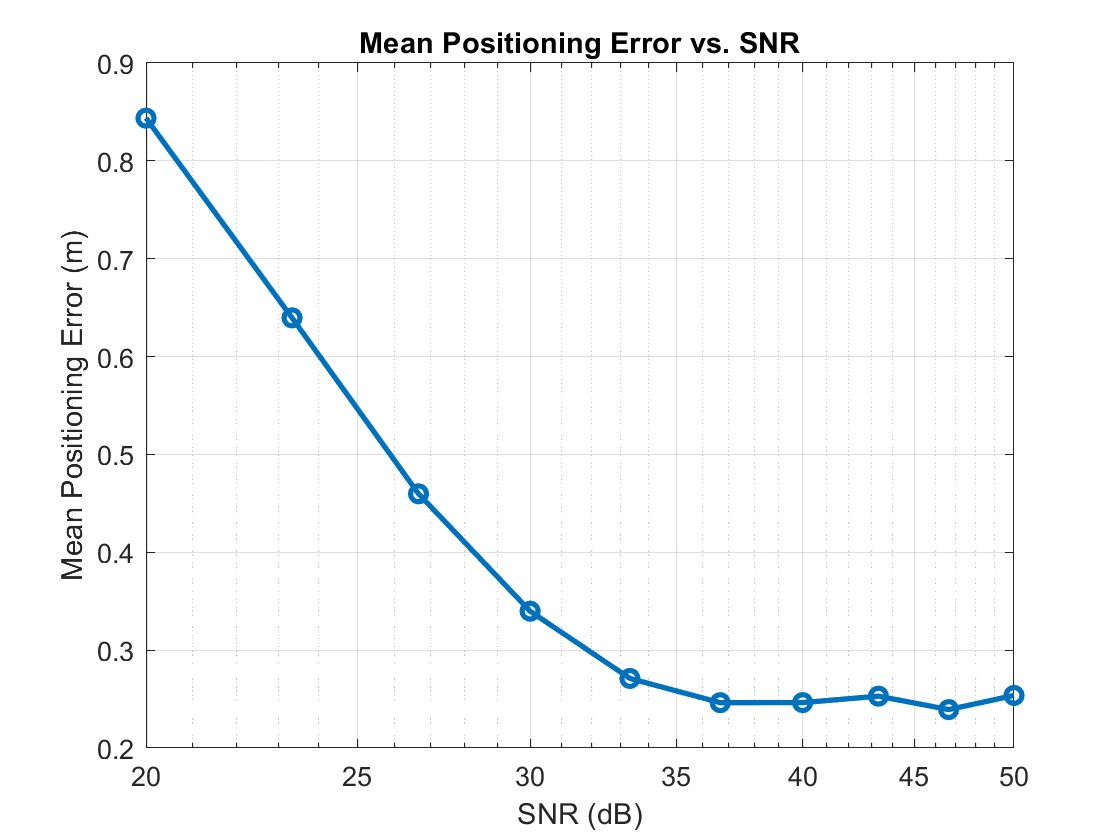}
    \caption{Mean positioning error vs \ac{snr} under random orientation.}
    \label{fig:mean error random}
\end{minipage}
\end{figure*}

In Fig. \ref{fig:cdf fix}, we plot the \ac{cdf} of the position errors of the proposed model under the assumption of fixed device orientation. As can be seen from the figure, the proposed system achieves high-precision localisation, with 95\% of the total 3D errors confined within 7cm. Notably, the X-axis and Y-axis exhibits the best performance, achieving sub-centimetre accuracy for over 80\% of the results. This is primarily due to the high accuracy of the scanning. In contrast, the Z-axis presents comparatively higher estimation errors, highlighting the system's sensitivity to vertical positioning due to geometry of the system. These results confirm the efficacy of the system for accurate indoor positioning in static scenarios, while also underscoring the importance of addressing vertical diversity and dynamic orientation variations in future enhancements.

The same is shown for the case of random user device orientation in Fig. \ref{fig:cdf rand}. However, the \ac{cdf} cannot show very clearly the difference between the random orientation and the fixed orientation, as shown in Fig. \ref{fig:cdf rand}, the shape of the figure is very identical to the fixed orientation except that for the random orientation, there exist about 8\% of sample points outside of the \ac{fov}. In order to better understand the effect of random orientation to the positioning accuracy, the following simulations are conducted.


As shown in Fig. \ref{fig:mean error fix} and Fig. \ref{fig:mean error random}, the analysis of mean positioning error at various grid points, with a space of 0.1m, as influenced by \ac{snr}, demonstrates notable performance variations in visible light positioning systems under both fixed and random orientation conditions. Under fixed receiver orientation, the mean positioning error exhibits a monotonic decrease as \ac{snr} increases from 20dB to 40dB. The error drops significantly from approximately 0.7m at 20dB to below 0.05m at 40dB, which decreases around 93\%, indicating a strong sensitivity of the positioning accuracy to \ac{snr} in the low-to-moderate \ac{snr} regime. It is noted that beyond 40dB, the error plateaus, stabilising around 0.02–0.03m, suggesting that the estimation algorithm reaches its performance limit, likely dictated by the system's geometric configuration or intrinsic modelling assumptions rather than measurement noise. This saturation implies that further improvements in \ac{snr} yield diminishing returns in terms of positioning accuracy.

In contrast, the random configuration demonstrates a similar general trend of decreasing error with increasing \ac{snr}, but with several key differences. First, the overall error is higher across the entire \ac{snr} range. At 20dB, the error starts at approximately 0.85m, and while it decreases with increasing \ac{snr}, the convergence behaviour is less favourable. From approximately 35dB onward, the error fluctuates in the range of 0.25–0.27m, without clear stabilisation or monotonicity. This saturation implies that orientation uncertainty introduces fundamental geometric ambiguities that cannot be resolved through signal enhancement alone. Therefore, while high \ac{snr} ensures robust performance in ideal static scenarios, the presence of orientation dynamics necessitates the integration of orientation estimation or diversity techniques to maintain positioning accuracy in practical environments.

\section{Conclusion}
In this paper, an indoor \ac{vlp} system with a single \ac{vcsel} deployed on the ceiling that employs a scanning facility is proposed. Using the received optical power as well as the incidence vector, the system provides three-dimensional user localisation employing a single transmitter. Simulation results confirmed that the technique could achieve high positioning accuracy with statically oriented receiver fixes such that errors decrease significantly with a higher \ac{snr}. However, if the receiver orientation was randomised, estimation error increased and started fluctuating randomly, thus orientation uncertainty remains one of the largest problems. The single-\ac{vcsel} solution provides a low-complexity and cost-effective approach for indoor positioning, which could be further enhanced by extending it to multi-user scenarios and real-time operation.

\balance
\bibliographystyle{IEEEtran}
\bibliography{References.bib}

\end{document}